\documentstyle[twoside,natbib,epsf]{article}

\input{ibvs2.sty}

\def\cite#1{\citealt{#1}}
\def\ibvs{Inf. Bull. Var. Stars}
\def\apj{ApJ}
\def\apjl{ApJ}
\def\iaucirc{IAU Circ.}
\def\mnras{MNRAS}

\begin{document}

\IBVShead{xxxx}{xx March 2002}

\IBVStitletl{V2540 O\lowercase{ph} (N\lowercase{ova} O\lowercase{ph} 2002):
            Large-Amplitude Slow Nova}\
            {with Strong Post-Outburst Oscillations}

\IBVSauth{Kato,~Taichi$^1$, Yamaoka,~Hitoshi$^2$, Ishioka,~Ryoko$^1$}
\vskip 5mm

\IBVSinst{Dept. of Astronomy, Kyoto University, Kyoto 606-8502, Japan, \\
          e-mail: (tkato,ishioka)@kusastro.kyoto-u.ac.jp}

\IBVSinst{Faculty of Science, Kyushu University, Fukuoka 810-8560, Japan,
          e-mail: yamaoka@rc.kyushu-u.ac.jp}

\IBVSobj{V2540 Oph}
\IBVStyp{N}
\IBVSkey{nova, astrometry, identification, classification}

\begintext

   V2540 Oph (Nova Oph 2002) was independently discovered by Katsumi Haseda
and Yuji Nakamura at magnitude 9.0 on 2002 January 24
(\cite{has02v2540ophiauc}).  \citet{ret02v2540ophiauc}
detected emission lines of hydrogen and Fe II, indicating that the object
is an Fe II class nova caught in the early decline stage.  Later examination
of photographs revealed that the nova was already at magnitude 8.9
on 2002 January 19 (\cite{sek02v2540ophiauc}).

   Since the detection of the outburst, the nova has been intensively
monitored by a number of observers. 
Figure 1 shows the light curve constructed from visual, CCD $V$-band and
photovisual observations reported to the VSNET Collaboration
(http://www.kusastro.kyoto-u.ac.jp/vsnet/).  The nova showed
strong post-maximum oscillations up to 1 mag.  The large-amplitude
early stage oscillations resemble those observed in V1178 Sco =
Nova Sco 2001) and V4361 Sgr = Nova Sgr 1996 (\cite{kat01v1178sco}).

   Superimposed on these oscillations, the nova showed a steady fade
at 0.033 mag d$^{-1}$.  [This rate was determined using
the data between 2452294 and 2452341, during which the general trend
of the fading can be approximated by a single decline rate.  The last part
of the light curve, when the nova underwent a long-lasting brightening,
was not used in this analysis.  If we incorporate the last part of the
light curve, the average decline rate becomes 0.013 mag d$^{-1}$, which
may more severely constrain the following discussion].
By applying the recently calibrated relation
(\cite{dellaval95novaabsmag}) of absolute maximum magnitude vs
rate-of-decline (MMRD) in classical novae, we obtain the expected
absolute $V$-band maximum magnitude of $M_{\rm V}$=$-$7.0$\pm$0.5.
We performed the accurate astrometry with the images obtained by 
Kyoto 0.30-m telescope taken on Mar. 8.23 UT, which revealed the 
position of the nova as: R.A. = 17$^{\rm h}$ 37$^{\rm m}$ 
34$^{\rm s}$.385 $\pm$ 0$^{\rm s}$.017, Decl. = $-$16$^\circ$ 
23$^\prime$ 18$^{\prime\prime}$.19 $\pm$ 0$^{\prime\prime}$.18 
(equinox 2000.0, using 59 UCAC1 reference stars).  This position is 
marginally consistent of the reported position by K. Kadota, who
measured Haseda's discovery films (\cite{has02v2540ophiauc}).
No corresponding object was found on 
DSS and 2MASS scans within 2$^{\prime\prime}$.5 of the nova, setting
an upper limit of the prenova magnitude of $\sim$21.

\IBVSfig{9cm}{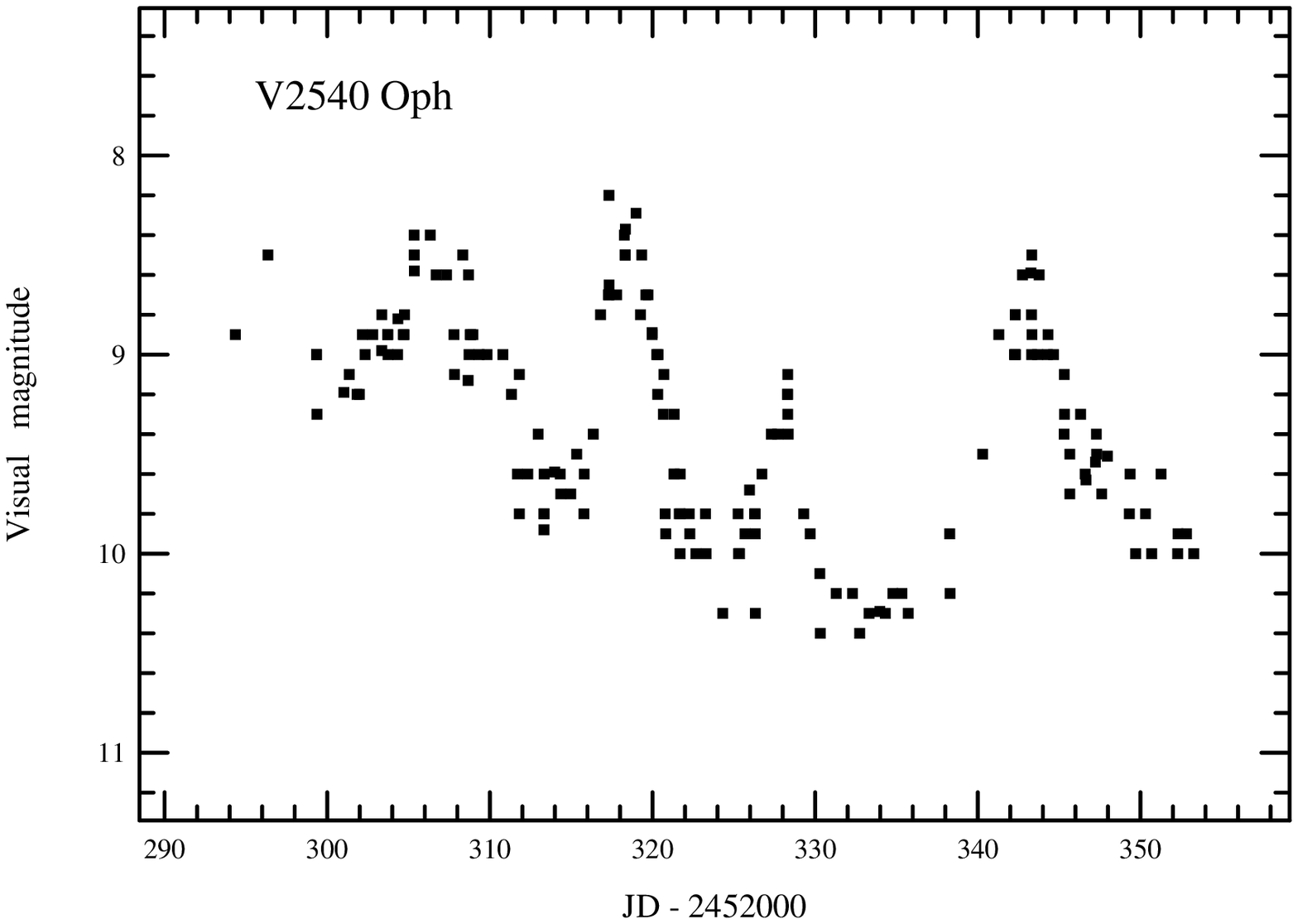}{
    Light curve of V2540 Oph (Nova Oph 2002) constructed from visual,
CCD $V$-band and photovisual observations reported to the VSNET
Collaboration.
}

\IBVSfig{10cm}{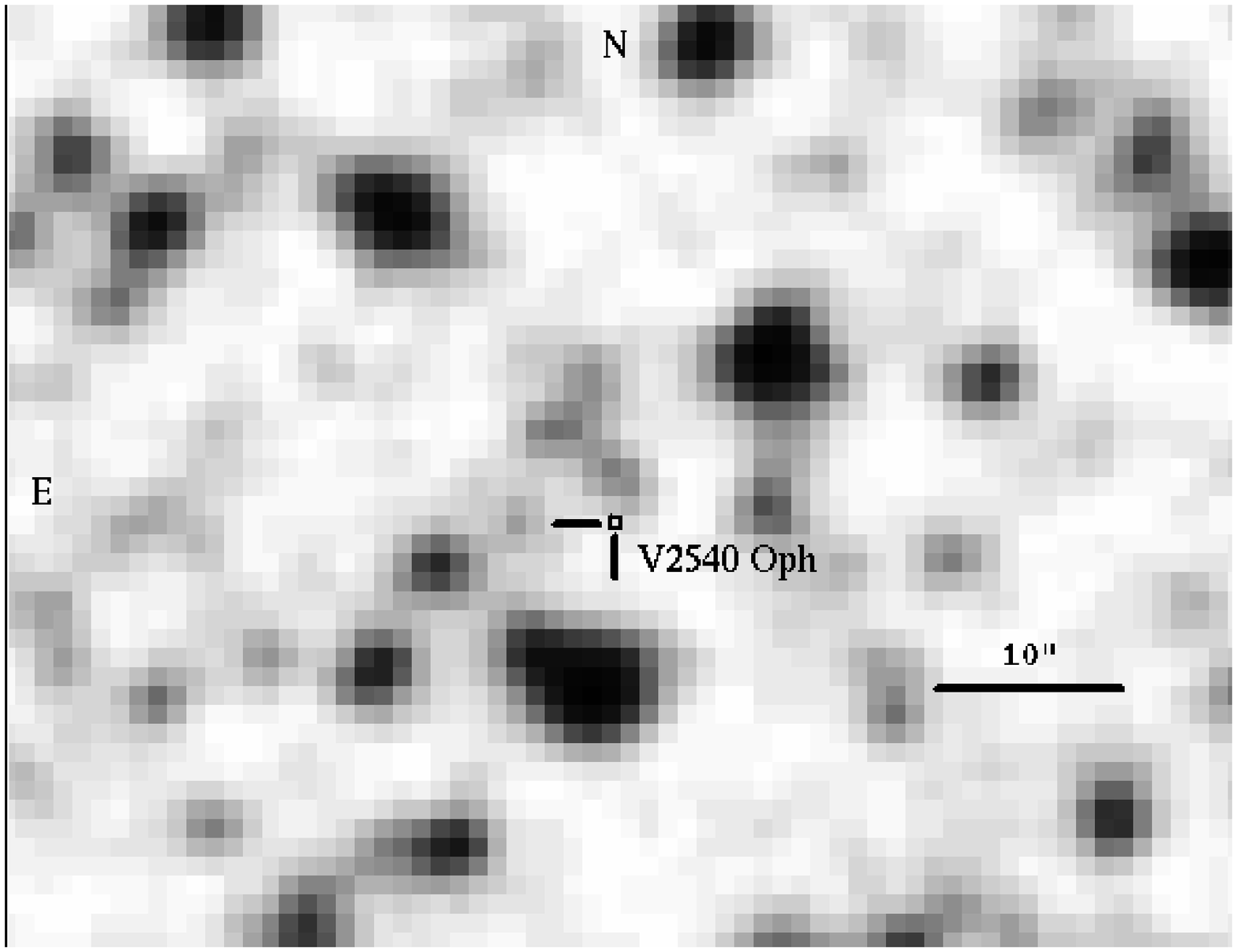}{
    The position of V2540 Oph (square) on DSS2 red image.  No prenova 
can be found to the image limit (mag $\sim$21).  The north is up, and 
the east is left.
}

This indicates that the lower limit of
the outburst amplitude is $\sim$12.5 (by adopting the observed maximum
magnitude of 8.5), which is unusually large for a slow nova
with a decay rate of 0.033 mag d$^{-1}$.
By using the above expected absolute $V$-band maximum magnitude, we can
set an upper limit of $M_{\rm V}$$\sim$5.5 for the nova progenitor.
This magnitude is extremely faint for known prenova magnitudes and other
novalike cataclysmic variables (\cite{war86NLabsmag,war87CVabsmag}).
[Available observations suggest that the true maximum of the
  nova msut have been missed.  By considering this, both the decline rate
  and the outburst amplitude could be larger than the values in this
  discussion.  However, we consider this effect will not severely affect
  the conclusion, because 1) the MMRD-relation (\cite{dellaval95novaabsmag})
  is known to be relatively flat (i.e. little depends on the decline rate)
  around the decline rate in question; a brighter maximum will therefore
  tend to pose a more stringent upper limit for the prenova), and 2)
  the reported spectrum (\cite{ret02v2540ophiauc}) suggests that the object
  was caught during an early decay stage.]

   Such a faint prenova magnitude would require a small mass-transfer
rate, a small dimension of the disk, or a high inclination.  Because
V2540 Oph is a slow nova, the low mass-transfer rate is a rather
unlikely explanation.  We propose that the nova should have either
a short orbital period or a high inclination.  We know an excellent
example of such a nova: V2214 Oph (Nova Oph 1988), which is believed
to be a magnetic nova inside the period gap (\cite{bap93v2214oph}).
V2540 Oph resembles V2214 Oph in many aspects: large outburst amplitude,
slow rate of decline, and the presence of prominent oscillations.
Since the characteristic double-wave orbital modulations were already
present during the decay stage of V2214 Oph (\cite{bap93v2214oph}),
we strongly encourage observers to detect orbital signatures in V2540 Oph.

\vskip 3mm

We are grateful to all observers who reported vital observations to VSNET.
This work is partly supported by a grant-in aid (13640239) from the
Japanese Ministry of Education, Culture, Sports, Science and Technology.
This research has made use of the 2MASS scan at NASA/IPAC Infrared Science
Archive, the Digitized Sky Survey producted by STScI, and the VizieR 
catalogue access tool.

\end{document}